\documentclass[aps,prd,preprint,tightenlines,
   showpacs,nofootinbib]{revtex4}
\usepackage{graphicx}
\usepackage{dcolumn}
\usepackage{bm}
\usepackage{amssymb} 
\newcommand{\beq}{\begin{equation}}
\newcommand{\eeq}{\end{equation}}
\newcommand{\beqa}{\begin{eqnarray}}
\newcommand{\eeqa}{\end{eqnarray}}

\newcommand{\vp}{\varphi}

\let\gam=w

\renewcommand{\epsilon}{\varepsilon}

\begin{document}


\vspace{8cm}

\title{Non-minimal scalar-tensor theories}

\author{
Ismael Ayuso$\,^{(a)}$\footnote{E-mail:isayuso@ucm.es},
and Jose A. R. Cembranos$\,^{(a,b,c)}$ \footnote{E-mail:cembra@fis.ucm.es},
}

\affiliation{$^{(a)}$ Departamento de F\'{\i}sica Te\'orica I, Universidad Complutense de Madrid, E-28040 Madrid, Spain;}
\affiliation{$^{(b)}$ Facultad de Ciencias, CUICBAS, Universidad de Colima, 28040 Colima, Mexico;}
\affiliation{$^{(c)}$ Dual C-P Institute of High Energy Physics, 28040 Colima, Mexico.}

\date{\today}

\vspace*{2cm}

\begin{abstract}
We compute the spectrum of scalar models with a general coupling to the scalar curvature.
We find that the perturbative states of these theories are given by two massive spin-0 modes in addition to one massless
spin-2 state. This latter mode can be identified with the standard graviton field. Indeed,
we are able to define an Einstein frame, where the dynamics of the massless spin-2 graviton is the one
associated with the Einstein-Hilbert action. We also explore the interactions of all these degrees of freedom
in the mentioned frame, since part of the coupling structure can be anticipated by geometrical arguments.
\end{abstract}

\pacs{04.50.-h, 04.60.-m, 98.80.Qc, 98.80.-k}


\maketitle


\section{Introduction}

There are different theoretical and experimental reasons to consider gravitational theories beyond General Relativity (GR) \cite{Mod_Grav_Theories}. Among them, we can mention
Lovelock theories \cite{Lovelock}, Gauss-Bonnet models \cite{GB}, extra dimensional geometrical extensions \cite{XD}, supergravity \cite{sugra},
non-local gravitational modifications \cite{NL}, Lorentz violating and CPT breaking models \cite{LV}, vector (or higher spin)-tensor models \cite{VT}, disformal gravity \cite{disformal} and Scalar-Tensor Theories (STTs) \cite{ST, COPU}. This latter case will be the particular subject of our work; i.e. we will study particular extensions of the gravitational interaction, that are defined by the addition of scalar degrees of freedom.
These new spin-0 states act as mediators of part of the total gravitational force.

Currently, there are only one candidate for a fundamental scalar particle.
Its discovery was announced in 2012 by both ATLAS \cite{:2012gk} and CMS \cite{:2012gu} collaborations.
It has associated a mass around 125 GeV and it is consistent with
the predictions for the so-called Higgs boson of the Standard Model (SM) of particles and interactions.
However, many other scalar fields are motivated by different theories, such as the
Jordan-Fierz-Brans-Dicke (JFBD) model \cite{stgen} or string theory \cite{polchy}, where
the scalar state is coupled to the matter sector through the trace of the energy-momentum tensor.
In such a case, thanks to a series of field redefinitions, it is possible to reexpress the extended
gravitational theory in terms of the Einstein gravity associated with General Relativity (GR)
with new couplings to the matter sector. A similar phenomenology is associated with the so-called
$f(R)$ theories \cite{fR,newnew,R2}, whose gravitational action is defined in terms of a given
function $f$, of the scalar curvature $R$. In addition, extended models with scalar fields non-minimally coupled
to gravity have been proposed in the present literature. In particular, the possibility of supporting
the early inflation with scalar models, which exhibits non-zero vacuum expectation values (VEVs) at low energies,
has inspired the Generalized Higgs Inflation Models (GHIMs) \cite{GHIM}.

As we have commented, the new interactions mediated by these scalar degrees of freedom can be interpreted
as part of the gravitational force. In addition, the phenomenology associated with these new degrees of freedom
themselves can be very rich and provide viable solutions to open problems in cosmology such as the mentioned inflation,
dark energy or dark matter \cite{R2}. However, these new models suffer important constraints. For instance,
this new field can lead to effective variations of fundamental constants, such as gauge and Yukawa couplings or
masses. In particular, precision tests of gravity, big bang nucleosynthesis (BBN)
\cite{speedup1,bbnJFBD,bbnst1,bbn_quad,dp,couv}, cosmic microwave
background anisotropies~\cite{ru02} and weak-lensing~\cite{carlo} constrain the phenomenology of these
scalar fields. In any case, different scalar models are not very sensitive to these restrictions
\cite{dnord,dpol,Khoury:2003aq,brax,newnew}.

The paper will be organized as follows.  In the
next section, we will define the Non-minimal Scalar Tensor Theory (NMSTT) which will be studied.
We will provide the general set of equations that defines the Einstein frame (EF)
through a conformal transformation. In Section III, we will report the general couplings,
which are enforced by such a transformation. They will be particularized for the
standard model (SM) content. We will then set up our formalism for treating different
gravitational theories in Section IV: JFBD model \cite{stgen}, $f(R)$ theories \cite{fR,newnew,R2} and
GHIMs \cite{GHIM}. In Section V, we will resume our main conclusions.

\section{Non-minimal scalar-tensor theories}
\label{cosmology-eqs}

The most general action for a scalar field non-minimally coupled to the scalar curvature $R$, associated with the space-time metric
$g_{\mu\nu}$, can be written in the following form:
\beq
S_{\text{NMSTT}}={1\over 2}\int d^4x\sqrt{-g}\left[-g^{\mu\nu}\nabla_{\mu}\vp\nabla_{\nu}\vp+\,J(\vp ,R)\right],
\label{AvpR}
\eeq
if we restrict the scalar field derivatives to the standard kinetic contribution (the first term). 
The function $J(\vp ,R)$ takes into account the mentioned coupling, that in general, it is non-separable.
Note that this term may include a potential (mass or self interaction) for the scalar field,
or a pure gravitational contribution without explicit dependence on $\vp$.
Unless otherwise specified, we will use reduced Planck units throughout this work
$(\kappa\equiv\sqrt{8\pi G}=c=\hbar=1)$, Greek indices run from 0 to 3, and the symbol $\nabla_{\mu}$ denotes the standard
covariant derivative defined with respect to the metric $g_{\mu\nu}$. 

In addition, we will assume a matter content, whose fields will be represented by the letter $\mu_i$, minimally
coupled to the space-time metric:
\beq
S_{\mu_i}=S_{i}(g_{\mu\nu},\mu_i),
\eeq
where by {\it matter}, we denote any field in the theory in addition to $\vp$ and $g_{\mu\nu}$.
In such a case, we can claim that we have defined the model in the Jordan Frame (JF), in which the metric $g_{\mu \nu}$
couples in this standard way to the matter content. Although the Action (\ref{AvpR}) did not
support explicitly a kinetic term for the gravitational interaction, this could be implicit due to the presence of the
Ricci scalar in the coupling $J(\vp ,R)$. In any case, the gravitational interaction between matter fields can suffer important
modifications. Within the JF, this effect seems natural to be interpreted as a modification of the Newton constant by the presence
of the scalar field $\vp$, but it means that the scalar mode is mediating part of the gravitational interaction. Finally,
we will show that the model given by Eq. (\ref{AvpR}) supports another scalar perturbative degree of freedom, that is also
coupled to the matter content and completes the gravitational force.

In order to clarify the spectrum of the model, it is convenient to work in the EF, that is defined by
a conformal transformation, which rewrites the action for the metric in the standard Einstein-Hilbert term. In fact, two
metrics ($g_{\mu\nu}^*$ and $g_{\mu\nu}$) are conformally related if there is a function $\Omega$, which verifies:
\beq
g_{\mu\nu}^*=\Omega^2g_{\mu\nu}\,.
\eeq
It implies that $g^{*{\mu\nu}}=\Omega^{-2}\,g^{\mu\nu}$, $\sqrt{-g}= \Omega^{-n} \sqrt{-g^*}$ and
\beq
R^*=\Omega^{-2}R-2(n-1)g_{*}^{\mu\nu}\,\nabla^{*}_{\mu}\nabla^{*}_{\nu}\ln\Omega
+(n-2)(n-1)g_{*}^{\mu\nu}\,\nabla^{*}_{\mu}\ln\Omega\,\nabla^{*}_{\nu}\ln\Omega\,,
\eeq
where $n$ is the dimension of the manifold, $R^*$ is the Ricci scalar associated with the metric $g_{\mu\nu}^*$, and $\nabla^{*}$
is its corresponding covariant derivative. The function $\Omega$ is dubbed the conformal factor of the transformation. We will make use of
the above expressions for $n=4$. In particular, for rewriting the action, it is particularly useful the following expression:
\beq
\sqrt{-g}R=
\Omega^{-2}\sqrt{-g^*}
[R^*+6g_*^{\mu\nu}\,\nabla^{*}_{\mu}\nabla^{*}_{\nu}\ln\Omega-6g_*^{\mu\nu}\,\nabla^{*}_{\mu}\ln\Omega\,\nabla^{*}_{\nu}\ln\Omega].
\eeq
In order to identify the correct conformal transformation that defines the EF, it is convenient to work with an auxiliary scalar field $\phi$, defined by the following equation:
\beq
J(\vp ,R)=J(\vp ,\phi)+J'(\vp ,\phi)(R-\phi)\,,
\eeq
where $'$ denotes the partial derivative of the $J$ function with respect to its second argument:
\beq
J'(\vp ,\phi)= \partial_\phi J(\vp,\phi) \,.
\eeq
We assume that $J''(\vp ,\phi)\neq 0$. We will discuss separately the case $J''(\vp ,\phi)= 0$ in following
sections, but we can advance that, in such a case, the auxiliary field cannot be defined.
By expressing $J(\vp ,R)$ in terms of $\phi$ in the Action (\ref{AvpR}), we can write a Lagrangian
that is linear on $R$. Indeed, the two actions are equivalent if we also impose optimization with respect
to $\phi$ in order to obtain the equations of motion, whose solution implies $\phi=R$.

Now, it is evident to identify the proper conformal factor associated with the Einstein metric, i.e. the metric
corresponding to the EF:
\beq
\Omega^2=J'(\vp,\phi)\,.
\eeq
where we are assuming explicitly $J'(\vp,\phi)> 0$. This sign is necessary to guarantee a positive Newton constant, i.e.
an attractive gravitational interaction mediated by the standard spin-2 massless graviton. In the opposite case,
the graviton has a {\it ghost} character since its kinetic term has the {\it wrong} sign.
Finally, we can define a new scalar $\Phi$, in terms of the fields $\vp$ and $\phi$:
\beq
\Phi=\sqrt{3/2}\,\ln J'(\vp,\phi)\,,
\label{scalars}
\eeq
so that, except for a boundary term, we can write the total action as:
\beq
S_{\text{NMSTT}}={1\over 2}\int d^4x \sqrt{-g^*}\left[
R^*
-g^{\mu\nu}_{*}\nabla^*_{\mu}\Phi\nabla^*_{\nu} \Phi
-g_*^{\mu\nu}{e^{-\sqrt{2\over{3}}\Phi}}\,\nabla^*_{\mu}\vp\nabla^*_{\nu}\vp
-2\,V(\vp,\Phi)
\right]\,,
\label{GJF}
\eeq
where:
\beq
V(\vp,\Phi)={1\over 2}\left[
\phi(\vp,\Phi) {e^{-\sqrt{2\over{3}}\Phi}}
- J(\vp,\phi(\vp,\Phi)) {e^{-2\sqrt{2\over{3}}\Phi}}
\right]\,.
\eeq
is the potential associated with the self-interaction of the scalar modes $\vp$ and $\Phi$, and
the interaction between them. One needs to use Eq. (\ref{scalars}) in order to write $\phi$
in terms of $\vp$ and $\Phi$. It is interesting to remark that the third term in Eq. (\ref{GJF})
does not only account for the standard kinetic term for $\vp$, but also for a derivative
interaction with $\Phi$. This follows, for example, by expanding the exponential factor around $\Phi=0$,
i.e. by assuming a small deviation from the two frames, a small difference between $g_{\mu\nu}$
and $g_{\mu\nu}^{*}$. On the other hand, the first term in (\ref{GJF}) corresponds to the standard
Einstein-Hilbert action and the second one is associated with a pure kinetic term for $\Phi$.
Of course, the kinetic terms include the interaction of both scalar fields with the geometry through the
metric tensor $g_{\mu\nu}^{*}$.

\section{Interaction with the matter content: the Standard Model}

It is interesting to analyze in more detail the interactions associated with the different
modes contained in the spectrum of the theory. Indeed, the coupling of the scalar field $\vp$
is open and not restricted by the geometrical structure of the model. The opposite situation
corresponds to the other scalar field $\Phi$ and the metric tensor. As we have discussed,
the matter content is explicitly coupled to $\Phi$ when the action is expressed in terms of
the Einstein metric:
\beq
S_{\mu_i}=S_{i}(e^{-\sqrt{2\over{3}}\Phi}\,g^*_{\mu\nu},\mu_i)\,.
\eeq
For example, we can follow \cite{R2} in order to detail the coupling of this scalar mode.
This computation can be done directly with the help of the original action (JF), but it is
more transparent and easier in the EF. We can study the couplings at the linear order 
by expanding perturbatively the Jordan metric over the Minkowski background \cite{R2}:
\begin{eqnarray}
\label{metricperturbation}
g_{\mu\nu}=\eta_{\mu\nu}+\frac{1}{2}h_{\mu\nu}^{*}
-\sqrt{\frac{2}{3}}\,\Phi\,\eta_{\mu\nu}\,,
\end{eqnarray}
where $h^*_{\mu\nu}$ takes into account the standard two degrees of freedom associated with
the spin-2 (traceless and divergence-free) graviton. For simplification, we assume a common Minkowski
background for the Jordan and Einstein geometries. In other words, we expand around $\Phi=0$. 
In such a case, for computing the linear order analysis,
we do not need to specify the frame for quantities such as the energy-momentum tensor.

By taking variations with respect to the metric in the matter action, it is evident that the spin-2
degree of freedom will have associated the standard interaction with the corresponding energy-momentum tensor.
On the other hand, the coupling of the spin-0 mode at the linear level will be given by the trace of the
same energy-momentum tensor:
\begin{eqnarray}
\label{coupling}
{\cal L}_{\Phi-T_{\mu\nu}}&=&\frac{1}{\sqrt{6}}\,\Phi\,T^{\mu}_{\;\;\mu}\,.
\end{eqnarray}
It means that $\Phi$ interacts with massive SM fields at tree level.
In particular, the three body couplings are given by:
\begin{eqnarray}
\label{mass}
{\cal L}^{\text{tree-level}}_{\Phi-\text{SM}} &=& \frac{1}{\sqrt{6}}\,\Phi\,
\left[2\,m_h^2 h^2-\nabla^*_\mu h\nabla_*^\mu h
+
\sum_\psi m_\psi\, {\bar \psi} \psi
 - 2\, m_W^2\, W^+_\mu {W^-}^\mu -\,m_Z^2\, Z_\mu Z^\mu
 \right]\,
\end{eqnarray}
with the Higgs boson ($h$), (Dirac) fermions ($\psi$), and electroweak gauge bosons ($W^\mu$ and $Z^\mu$),
respectively. In addition, this scalar field interacts with photons and gluons by radiative corrections
induced at one loop by charged gauge bosons and fermions (i.e. due to the {\it conformal anomaly} \cite{R2}):
\begin{eqnarray}
\label{massless}
{\cal L}^{\text{one-loop}}_{\Phi-\text{SM}} &=&
\frac{1}{\sqrt{6}}\,\Phi\,
\left[
{\alpha_{\text{EM}} c_{\text{EM}} \over  8 \pi}\, F_{\mu\nu} F^{\mu\nu}
+
{\alpha_{\text{s}} c_{\text{G}}  \over 8\pi}\, G^a_{\mu\nu} G_a^{\mu\nu}
\right]\,,
\end{eqnarray}
where $F_{\mu\nu}$ is the guage invariant electromagnetic field strength tensor,
$G^a_{\mu\nu}$ represents the gluon field strength tensor,
$\alpha_{\text{EM}}$ is the fine-structure constant, and
$\alpha_{s}$ is the strong coupling constant.
The particular value of the couplings $c_{\text{EM}}$ and $c_{\text{G}}$, depends on the energy
and the complete set of particles charged with respect to these gauge interactions.

\section{Scalar spectrum}

Another general property given by the geometrical structure of the model is that the kinetic term for
$\vp$ has the same coupling with $\Phi$ as the
matter fields. Indeed, $\Phi$ can be understood as a dilaton, that parameterizes the conformal factor
and couples to the trace of the energy-momentum tensor. However, there is a general mixing of the scalar
sector of the theory trough the mass matrix that is defined by the potential function $V(\vp,\Phi)$.
If we assume that this potential reaches a minimum of value $V_0$ at $(\vp_0,\Phi_0)$ (i.e. $V(\vp_0,\Phi_0)=V_0$),
the Squared-Mass Matrix (SMM) is given by:
\beq
M_{\vp\,\Phi}^2=\left.\left(
      \begin{array}{cc}
       e^{\sqrt{2\over{3}}\Phi_0}\; \partial^2_{\vp\,\vp}  V(\vp,\Phi) \;\; & \;\; 
       e^{\sqrt{1\over{6}}\Phi_0}\; \partial^2_{\vp\,\Phi}  V(\vp,\Phi)  \\
       e^{\sqrt{1\over{6}}\Phi_0}\; \partial^2_{\Phi\,\vp} V(\vp,\Phi) \;\; & \;\; 
                                    \partial^2_{\Phi\,\Phi} V(\vp,\Phi)  \\
      \end{array}
    \right)\right|_{(\vp_0,\Phi_0)}\,.
\eeq
Therefore, the mass eigenstates cannot be generally identified either with $\vp$ or with $\Phi$,
but with a linear mixing of both. This fact is not in contradiction with the dilaton nature
of the couplings associated with $\Phi$. The question is that the SMM breaks,
in general, scale invariance explicitly.

In any case, a non trivial mixing and even the presence of two scalar degrees of freedom is not completely general.
There are particular forms or values of the function $J(\vp,\phi)$ that are interesting to analyze separately as
we do in the following paragraphs.

\subsection{Linear Couplings}

In the particular case of a linear coupling of the field $\vp$ with the Ricci scalar, the Action (\ref{AvpR})
has associated a truncated scalar spectrum. Indeed, in such a case, we can write $J(\vp,R)$ in terms of
two functions of $\vp$: the one that parameterizes its non-minimal interaction $H(\vp)$, and the one that
defines its potential $U(\vp)$:
\beq
J(\vp,R) = H(\vp)\,R-2U(\vp)\,.
\eeq
In this case, it is not necessary to introduce any additional scalar field, since it is possible to define a
conformal transformation to the EF, through a conformal factor that depends only on
$\vp$, namely:
\beq
\Omega^2=H(\vp).
\eeq
Following for example \cite{COPU}, if we redefine the scalar field in the following way:
\beq
{1\over{2H(\vp)}}={1\over{2}}\left({d\vp_*\over{d\vp}}\right)^2-{3\over{4}}\left({d\ln H(\vp)\over{d\vp}}\right)^2\,,
\eeq
we can write the action in the EF with a standard kinetic term for $\vp_*$:
\beq
S_{\text{NMSTT}}=\int d^4x\sqrt{-g^*}
\left[{1\over 2}R^*
-{1\over 2}g_*^{\mu\nu}\nabla^*_{\mu}\vp_*\nabla^*_{\nu}\vp_*
-V(\vp_*)\right],
\eeq
where the potential for $\vp_*$ takes into account the conformal factor and the field redefinition:
\beq
V(\vp_*)={U(\vp(\vp_*))\over{H^2(\vp(\vp_*))}}\,.
\eeq
By assuming a minimum for this potential at $\vp_*^0$, the corresponding squared-mass of the scalar mode
will be given by:
\beq
m_{\vp_*}^2=\partial^2_{\vp_* \vp_*}V(\vp_*)\,|_{\vp_*^0}\,.
\eeq
This simple example illustrates the complexity in the identification of the scalar states. In this case, it is the
field $\vp$ (or $\vp_*$), the one that has associated the dilaton couplings with the matter content,
since the conformal transformation is parameterized by $\vp$ (or $\vp_*$):
\beq
S_{\mu_i}=S_{i}\left({g^*_{\mu\nu}\over{H(\vp(\vp_*))}},\mu_i\right)\,.
\eeq

\subsection{$f(R)$ theories}

Other simple examples of models described by Eq. (\ref{AvpR}) are the so-called $f(R)$ theories.
In this case, the $J(\vp,R)$ can be written as the sum of two functions, one depending on $R$ and another one
depending on $\vp$:
\beq
J(\vp,R) = f(R)-2U(\vp)\,.
\eeq
The first one gives the name to these models and the second one constitutes a standard potential for the
scalar field $\vp$. In this case, $\vp$ is minimally coupled to gravity, and it can be interpreted as part
of the matter content. However, the non-linear dependence on $R$ introduces an additional scalar degree of freedom.
Here, we can just particularize the equations derived in the previous section for the general case.
In fact, the new scalar field, when properly normalized, is defined by
\beq
\Phi=\sqrt{3/2}\,\ln f'(\phi)\,,
\label{fscalars}
\eeq
where the auxiliary field $\phi$, verifies $\phi=R$ by taking into account the equations of motion.
Here, $'$ means the derivative with respect to the unique argument $\phi$. The corresponding conformal factor
is:
\beq
\Omega^2=f'(\phi).
\eeq
Note that the total potential for the scalar sector cannot be written in general as the sum
of two individual potentials associated with each one of the fields:
\beq
V(\vp,\Phi)={1\over 2}\left[
\phi(\Phi) {e^{-\sqrt{2\over{3}}\Phi}}
- f(\phi(\Phi)) {e^{-2\sqrt{2\over{3}}\Phi}}
+2 U(\vp) {e^{-2\sqrt{2\over{3}}\Phi}}
\right]\,.
\eeq
The reason is that the conformal factor introduces a non-derivative interaction between the two scalar modes.
Therefore, even in this case, the off-diagonal entries of the SMM are not necessarily zero and the
mass eigenstates cannot be identified, in general, with $\vp$ or with $\Phi$.

\subsection{Generalized Higgs inflation models}

Non-minimal gravitational couplings of the SM Higgs doublet have been considered in order to
build viable models of inflation in the early universe \cite{HI}. Generalizations of this idea
with non-linear couplings to the Ricci scalar have been discussed in the literature with the SM Higgs working
as inflaton or with a similar scalar field \cite{GHIM}. As far as we know, the presence of a new scalar degree of freedom
has been missed in these analyses. As we have shown, the non-linear couplings in the Ricci
scalar introduce a new degree of freedom. We can determine its phenomenology
by using the general equations deduced in this work. The general $J(\vp,R)$ function that defines these
models can be written as
\beq
J(\vp,R) = R+\xi\,\vp^a\,R^b-2U(\vp)\,,
\eeq
where the first term is associated with the initial Einstein-Hilbert action for the Jordan metric; the second term is the non-linear coupling
parameterized by the strength constant $\xi$ and the exponents $a$ and $b$; and $U(\vp)$ is the potential for the
scalar field in the JF. Note that $a$ and $b$ need to be integer numbers to have an analytical interaction at
$\vp=0$ and $R=0$ respectively, but the non-minimal coupling can be also defined for any real value of both exponents.
The standard potential in these models is usually assumed to be:
\beq
U(\vp)={\lambda\over{4}}
\left(
\vp^2-{\mu^2\over{\lambda}}
\right)^2
\,,
\label{U}
\eeq
with $\,\mu,\,\lambda>0$. It implies that $U(\vp)$ is bounded from below, and develops a stable minimum at
$\vp_0=\mu/\sqrt{\lambda}$ with $U(\vp_0)=0$, i.e. we have
avoided the introduction of a vacuum energy\footnote{The potential develops an analogous minimum at
$\vp_0=-\mu/\sqrt{\lambda}$. The same discussion applies when the system chooses this other vacuum state.}.
By taking into account our previous results,
the conformal transformation to the EF is defined by
\beq
\Omega^2=1+\xi\,b\,\vp^a\,\phi^{b-1}\,,
\eeq
with $\phi=R$ as we have commented. Therefore, provided $b\neq 1$, the coupling introduces a new scalar
particle, associated with the normalized field:
\beq
\Phi=\sqrt{3/2}\,\ln (1+\xi\,b\,\vp^a\,\phi^{b-1})\,.
\label{Hscalars}
\eeq
As we have discussed, this new degree of freedom is associated with the dynamics of the Jordan Ricci scalar
through the relation: $R=\phi=\{[\exp(\sqrt{2/3}\,\Phi)-1]/(\xi\,b\,\vp^a)\}^{1/(b-1)}$.
Therefore, the total action for this type of GHIMs is writen in the EF as:
\beq
S_{\text{GHIM}}={1\over 2}\int d^4x \sqrt{-g^*}\left[
R^*
-g^{\mu\nu}_{*}\nabla^*_{\mu}\Phi\nabla^*_{\nu} \Phi
-g_*^{\mu\nu}{e^{-\sqrt{2\over{3}}\Phi}}\,\nabla^*_{\mu}\vp\nabla^*_{\nu}\vp
-2\,V(\vp,\Phi)
\right]\,,
\eeq
where:
%
\begin{eqnarray}
V(\vp,\Phi)&=&{1\over 2}
\left[
\xi\vp^a (b-1)
\left(
\frac{e^{\sqrt{2\over{3}}\Phi}-1}{\xi\,b\,\vp^a}
\right)^{b\over{b-1}}
+2U(\vp)
\right]
{e^{-2\sqrt{2\over{3}}\Phi}}\,.
\end{eqnarray}
In order to simplify the discussion, we can fix the particular values: $a=b=2$, and $\xi>0$. In such a case, the total potential is bounded from below provided $U(\vp)$ is it as well.
Indeed, a minimum of the scalar sector can be found at $(\vp_0,\Phi_0)=(\mu/\sqrt{\lambda},0)$ if the Potential (\ref{U}) is assumed. In fact,
such a minimum is global since $V(\vp_0,\Phi_0)=0$ and $V(\vp,\Phi)$ is non-negative.
In this case, the SMM around the minimum is given by:
\beq
M^2_{\vp\Phi}
    =\left(
      \begin{array}{cc}
        2 \mu^2 & 0                                \\
        0       & \frac{\lambda}{6\xi\mu^2}  \\
      \end{array}
    \right)
    \,.
\eeq
Therefore, in this particular case, the mass eigenstates can be identified with $\vp$ and $\Phi$. In other words,
they do not mix. In addition, the phenomenology of $\Phi$ is decoupled since the mass scale $\mu$ is expected
to be very small with respect to the Planck scale, what implies a very large mass for $\Phi$
(by assuming $\lambda,\xi \sim 1$).

It is important to comment that the original action has a parity symmetry associated with the sign of the scalar
$\vp$. However this is not the fundamental reason for the non-mixing of the scalar fields, since this discrete
$\mathbb{Z}_2$ symmetry is broken by the VEV of $\vp$. It means that this symmetry will not be able to protect this property, that will be potentially destroyed by radiative corrections.

Coming back to the {\it on-shell} analysis, the situation is different for $b=2$ and $a<-2$. In this case,
$\Phi$ is lighter than $\vp$ for $\mu\ll 1$ ($\lambda,\xi \sim 1$). Indeed, for such values of the exponents, the SMM is still diagonal, but the
non-zero entries are $M^2_{\vp\vp}=2 \mu^2$ and $M^2_{\Phi\Phi}=\lambda^{a/2}/(6\xi\mu^a)$. Finally, we must mention
that any other integer value of $b$ ($b>2$) may imply strong instabilities for the field configuration defined by
$(\vp_0,\Phi_0)=(\mu/\sqrt{\lambda},0)$, since the potential will develop a singularity at $\Phi=0$. Indeed, the
study of the SMM deduced in this work within the EF, is the most efficient way to analyze the stability of a NMSTT
(in the same way as for $f(R)$ theories, as it was originally pointed out in \cite{newnew}). This procedure is
equivalent to the Hessian matrix analysis of an optimization study in two variables.

\section{Discussion}

\label{discussion}

In this work, we have studied the phenomenology of NMSTTs, i.e. scalar field models defined by a general coupling of the scalar field with the
Ricci scalar. These theories can be understood as generalizations of the gravitational interaction written in a particular JF.
For the first time, we have found the EF corresponding to such general theories. We have characterized the conformal transformation that defines the relation between both frames. By following the general set of equations associated with the transformation between the two corresponding metric tensors, it is explicit that the spectrum of the theory contains not one, but two (generally massive) scalar degrees of freedom, in addition
to a massless spin-2 state. The latter particle can be associated with the standard mediator of Einstein gravity, whereas the second scalar mode is
related to the non-linear coupling of the original (JF defined) scalar field to the Ricci scalar. Indeed, provided
that the coupling is linear, we have proved that the general conformal transformation is not well defined and the spectrum of the theory is truncated by removing the second scalar degree of freedom. The situation is more involved if the Jacobian associated with the field redefinition is zero for particular values of the fields. In such a case, the effective number of degrees of freedom of the theory depends on the field configuration. This fact can be understood as if the spectrum of the theory maximizes the number of perturbative states, but the masses of the scalar modes depend on the values of the fields and can diverge for one of the modes. This fact removes effectively one of the scalar degrees of freedom as it was discussed in \cite{newnew,R2} for the case of $f(R)$ theories.

Once the degrees of freedom were identified, we have analyzed their couplings with the matter content.
In particular, we have studied the
couplings with SM particles. Under general assumptions, the spin-2 state couples as the standard GR graviton. The scalar degree of freedom associated with the conformal factor couples through the trace of the energy-momentum tensor. Indeed, it can be identified with a dilaton since it parameterizes general scale transformations. On the other hand, the coupling of the other spin-0 particle is completely model dependent. It changes by depending
on the definition of its interactions in the original action (JF), that is not fixed. It is interesting to remark that this
factorization of the couplings associated with the scalar content of the theory is simple because it is discussed in terms of the interaction eigenstates. These modes are not necessarily the mass eigenstates. In general, there is a mixing between the two scalar modes that leads to the rich phenomenology associated with these NMSTTs.

\vspace{.2 cm}

\section*{Acknowledgements}
We are thankful to J. Beltran Jimenez and A. de la Cruz-Dombriz for very useful comments.
This work has been supported by MICINN 
(Spain) project numbers FIS2011-23000, FPA2011-27853-C02-01, and
the Consolider-Ingenio MULTIDARK CSD2009-00064.


\begin{references}

\bibitem{Mod_Grav_Theories}
  A. Dobado and A. L. Maroto, Phys. Lett. B \textbf{316}, 250, (1993) [Erratum-ibid. B \textbf{321}, 435, (1994)];
  S.~Nojiri and S.~D.~Odintsov, eConf {\bf C0602061}, 06 (2006)   [Int.\ J.\ Geom.\ Meth.\ Mod.\ Phys.\  {\bf 4}, 115 (2007)];
  Phys.\ Rept.\  {\bf 505}, 59 (2011); 
  S.~Capozziello and M.~Francaviglia, Gen.\ Rel.\ Grav.\  {\bf 40}, 357 (2008); 
  T.~P.~Sotiriou and V.~Faraoni, Rev. Mod. Phys. \textbf{82} 451 (2010); 
  F.~S.~N.~Lobo, arXiv:0807.1640 [gr-qc]; 
  S. Capozziello  and  V. Faraoni, {\it Beyond Einstein Gravity}, Fundamental Theories of Physics Vol. 170, Springer Ed., Dordrecht  (2011).

\bibitem{Lovelock}
  C.~Lanczos, Z. Phys. {\bf 73}, 147, (1932); Annals Math. {\bf 39}, 842, (1938);
  D.~Lovelock, J.\ Math.\ Phys.\  {\bf 12}, 498 (1971).

\bibitem{GB}
  G. Cognola, E. Elizade, S. Nojiri, S. D. Odintsov and S. Zerbini, Phys. Rev. D \textbf{73} 084007 (2006); 
  S.~Nojiri, S.~D.~Odintsov, Phys. Lett. B \textbf{631} 1 (2005); 
  Phys. Rev. D \textbf{68}, 123512 (2003); 
  E.~Elizalde, R.~Myrzakulov, V.~V.~Obukhov and D.~S\'aez-G\'omez, Class. Quant. Grav. \textbf{27}  095007 (2010); 
  R.~Myrzakulov, D.~S\'aez-G\'omez and A.~Tureanu, Gen.\ Rel.\ Grav.\ \ {\bf 43} 1671 (2011); 
  A. de la Cruz-Dombriz and D. S\'aez-G\'omez, Class. Quantum Grav. {\bf 29}  245014, (2012). 

\bibitem{XD}
  J.~Alcaraz {\it et al.}, Phys. Rev. D {\bf 67}, 075010 (2003); 
  P. Achard {\it et al.}, Phys. Lett. B {\bf 597}, 145 (2004); 
  J.~A.~R.~Cembranos, A.~Dobado and A.~L.~Maroto,  Phys.\ Rev.\ Lett.\  {\bf 90}, 241301 (2003); 
  Phys.\ Rev.\ D {\bf 68}, 103505 (2003); 
  AIP Conf.Proc. {\bf 670}, 235 (2003); 
  Int. J. Mod. Phys. D {\bf 13}, 2275 (2004); 
  Phys. Rev. D {\bf 70}, 096001 (2004); 
  Phys.\ Rev.\ D {\bf 73}, 035008 (2006); 
  Phys.\ Rev.\ D {\bf 73}, 057303 (2006); 
  J.\ Phys.\ A  {\bf 40}, 6631 (2007); 
  J.~A.~R.~Cembranos  {\it et al.},
  Phys.\ Rev.\ D {\bf 86}, 103506 (2012); 
  JCAP {\bf 1304}, 051 (2013); 
  JHEP {\bf 1309}, 077 (2013); 
  Phys.\ Rev.\ D {\bf 88}, 075021 (2013). 

\bibitem{sugra}
  D.~Z.~Freedman, P.~van Nieuwenhuizen and S.~Ferrara, Phys.\ Rev.\ D {\bf 13}, 3214 (1976); 
  S.~Deser and B.~Zumino, Phys.\ Lett.\ B {\bf 62}, 335 (1976);  
  E.~Cremmer, B.~Julia and J.~Scherk, Phys.\ Lett.\ B {\bf 76}, 409 (1978); 
  L.~J.~Hall, J.~D.~Lykken and S.~Weinberg, Phys.\ Rev.\ D {\bf 27}, 2359 (1983); 
  N.~Ohta, Prog.\ Theor.\ Phys.\  {\bf 70}, 542 (1983); 
  L.~Alvarez-Gaume, J.~Polchinski and M.~B.~Wise,  Nucl.\ Phys.\ B {\bf 221}, 495 (1983); 
  H.~P.~Nilles, Phys.\ Rept.\  {\bf 110}, 1 (1984). 
  J.~A.~R.~Cembranos, J.~L.~Feng, A.~Rajaraman and F.~Takayama, Phys.\ Rev.\ Lett.\  {\bf 95}, 181301 (2005); 
  AIP Conf.\ Proc.\  {\bf 903}, 591 (2007). 
  J.~A.~R.~Cembranos, J.~L.~Feng and L.~E.~Strigari, Phys.\ Rev.\ Lett.\  {\bf 99}, 191301 (2007); 
  Phys.\ Rev.\  D {\bf 75}, 036004 (2007); 
  M.~R.~Garousi, arXiv:1210.4379 [hep-th].

\bibitem{NL}
  T.~Biswas, T.~Koivisto and A.~Mazumdar, JCAP {\bf 1011}, 008 (2010); 
  T.~Biswas {\it et al.}, Phys.\ Rev.\ Lett.\  {\bf 104}, 021601 (2010); 
  JHEP {\bf 1010}, 048 (2010); 
  Phys.\ Rev.\  D {\bf 82}, 085028 (2010). 
  T.~Biswas, E.~Gerwick, T.~Koivisto and A.~Mazumdar, Phys.\ Rev.\ Lett.\  {\bf 108}, 031101 (2012); 
  T.~Biswas, A.~S.~Koshelev, A.~Mazumdar and S.~Y.~.Vernov, JCAP {\bf 1208}, 024 (2012). 

\bibitem{LV}
  V. A. Kostelecky and S. Samuel, Phys. Rev. D {\bf 39}, 683 (1989).
  D. Colladay and V. A. Kostelecky, Phys. Rev. D {\bf 55}, 6760 (1997);
  J. R. Ellis, N. E. Mavromatos and D. V. Nanopoulos,  Phys. Rev. D {\bf 61}, 027503 (1999);
  J. Alfaro, H. A. Morales-Tecotl and L. F. Urrutia, Phys. Rev. Lett. {\bf 84}, 2318 (2000);
  G. Amelino-Camelia, Nature {\bf 418}, 34 (2002);
  J. Magueijo and L. Smolin, Phys. Rev. Lett. {\bf 88}, 190403 (2002);
  J.~A.~R.~Cembranos, A.~Rajaraman and F.~Takayama, hep-ph/0512020; 
  Europhys.\ Lett.\  {\bf 82}, 21001 (2008); 
  S. Ghosh and P. Pal, Phys. Rev. D {\bf 75}, 105021 (2007).


\bibitem{VT}
  L.~H.~Ford, Phys.\ Rev.\ D {\bf 40} (1989) 967.
  J.~Beltr\'an Jim\'enez and A.~L.~Maroto, Phys.\ Rev.\ D {\bf 78} (2008) 063005; JCAP {\bf 0903} (2009) 016;
  Phys.\ Rev.\ D {\bf 80} (2009) 063512;
  T.~Koivisto and D.~F.~Mota, JCAP {\bf 0808}, 021 (2008); 
  J.~A.~R.~Cembranos  {\it et al.}, Phys.\ Rev.\ D {\bf 86}, 021301 (2012); 
  Phys.\ Rev.\ D {\bf 87}, no. 4, 043523 (2013); 
  JCAP {\bf 1403}, 042 (2014). 

\bibitem{disformal}
  G.~W.~Horndeski, Int.\ J.\ Theor.\ Phys.\  {\bf 10}, 363 (1974);  
  J.~D.~Bekenstein, Phys.\ Rev.\ D {\bf 48}, 3641 (1993); 
  J.~A.~R.~Cembranos {\it et al.}, Phys. Rev. D {\bf 65} 026005 (2002); 
  JCAP {\bf 0810}, 039 (2008); 
  Phys.\ Rev.\  D {\bf 83}, 083507 (2011); 
  Phys.\ Rev.\ D {\bf 84}, 083522 (2011); 
  Phys.\ Rev.\ D {\bf 85}, 043505 (2012); 
  J.~A.~R.~Cembranos and L.~E.~Strigari, Phys.\ Rev.\  D {\bf 77}, 123519 (2008); 
  M.~Zumalacarregui, T.~S.~Koivisto, D.~F.~Mota and P.~Ruiz-Lapuente, JCAP {\bf 1005}, 038 (2010); 
  T.~S.~Koivisto, D.~F.~Mota and M.~Zumalacarregui, arXiv:1205.3167 [astro-ph.CO]; 
  arXiv:1210.8016 [astro-ph.CO]. 

\bibitem{ST}
  C. Brans and R. H. Dicke, Phys. Rev., {\bf 124} 925 (1961);
  C. H. Brans, Phys. Rev., {\bf 125(6)} 2194 (1962);
  J. Garc\'ia-Bellido, A. Linde and D. Linde, Phys. Rev. D {\bf 50}, 730 (1994);
  J.~A.~R.~Cembranos, A.~de la Cruz Dombriz and L.~O.~Garcia, Phys.\ Rev.\ D {\bf 88}, 123507 (2013).

\bibitem{COPU}
  J.~A.~R.~Cembranos  {\it et al.}, JCAP {\bf 0907}, 025 (2009). 

\bibitem{:2012gk}
  G.~Aad {\it et al.}  [ATLAS Collaboration],
  Phys.\ Lett.\ B {\bf 716}, 1 (2012)
  [arXiv:1207.7214 [hep-ex]].

\bibitem{:2012gu}
  S.~Chatrchyan {\it et al.}  [CMS Collaboration],
  Phys.\ Lett.\ B {\bf 716}, 30 (2012)
  [arXiv:1207.7235 [hep-ex]].

\bibitem{stgen}
 P. Jordan, Nature (London) {\bf 164}, 637 (1949);
 M. Fierz, Helv. Phys. Acta {\bf29}, 128 (1956);
 C. Brans and R. Dicke, Phys. Rev. {\bf124}, 925 (1961);
 P. G. Bergmann, Int. J. Theor. Phys. {\bf 1}, 25 (1968);
 K. Nordtvedt, Astrophys. J. {\bf 161}, 1059 (1970);
 R. Wagoner, Phys. Rev. D{\bf 1}, 3209 (1970).

\bibitem{polchy}
 J. Polchinsky, {\it String theory}
 (Cambridge University Press, 1998).

\bibitem{fR}
K.~S.~Stelle, {\em Gen. Rel. Grav.} {\bf 9} 353 (1978);
A. A. Starobinsky, {\em Phys. Lett.} {\bf B 91}, 99 (1980);
M.~B.~Mijíc, M.~S.~Morris and W.~M.~Suen, Phys.\ Rev.\  D {\bf 34} 2934-2946 (1986);
S. M. Carroll, V. Duvvuri, M. Trodden and M. S. Turner \prd {\bf 70} 043528 (2004);
N. Goheer, J. Larena, P. K. S Dunsby, {\em Phys. Rev.} D {\bf 80} 061301 (2009);
S. Carloni, P. K. S. Dunsby, A. Troisi, {\em Phys. Rev.} D {\bf 77} 024024 (2008);
K. N. Ananda, S. Carloni, P. K. S. Dunsby, {\em Phys. Rev.} D {\bf 77}, 024033 (2008);  {\em Class. Quant. Grav.} {\bf 26} 235018 (2009);
J. A. R. Cembranos {\it et al.}, AIP Conf. Proc. {\bf 1458}  491 (2011); arXiv:1109.4519 [gr-qc]; 
A.~Abebe, M.~Abdelwahab, A.~de la Cruz-Dombriz and P.~K.~S.~Dunsby, {\em Class.\ Quant.\ Grav.}\  {\bf 29}, 135011 (2012);
A.~Abebe, A.~de la Cruz-Dombriz and P.~K.~S.~Dunsby, {\em Phys. Rev.} D {\bf 88} 044050 (2013). 
A. de la Cruz-Dombriz,  P. K. S. Dunsby, V. C. Busti,  S. Kandhai, arXiv:1312.2022 [gr-qc].

\bibitem{newnew}
  J.~A.~R.~Cembranos, Phys.\ Rev.\ D {\bf 73}, 064029 (2006). 

\bibitem{R2}
J.~A.~R.~Cembranos, Phys.\ Rev.\ Lett.\  {\bf 102}, 141301 (2009). 

\bibitem{GHIM}
  M.~Atkins and X.~Calmet, Phys.\ Lett.\ B {\bf 697}, 37 (2011); 
  G.~Chakravarty, S.~Mohanty and N.~K.~Singh, Int.\ J.\ Mod.\ Phys.\ D {\bf 23}, no. 4, 1450029 (2014);
  I. Oda, Phys. Rev. {\bf D 87}, 065025 (2013); 
  Phys. Lett. {\bf B 724}, 160 (2013); 
  Adv. Studies in Theor. Phys. {\bf 8}, 215 (2014); 
  R. Kallosh and A. Linde, JCAP {\bf 1306}, 027 (2013); 
  JCAP {\bf 1306}, 028 (2013); 
  JCAP {\bf 1307}, 002 (2013); 
  I. Bars, P. Steinhardt and N. Turok, Phys. Lett. {\bf B 726}, 50 (2013); 
  Phys. Rev. {\bf D 89}, 043515 (2014); 
  M. P. Hertzberg, arXiv:1403.5253 [hep-th];
  R. Costa and H. Nastase, arXiv:1403.7157 [hep-th].
  J.~Ren, Z.~Z.~Xianyu and H.~J.~He, JCAP {\bf 1406}, 032 (2014);
  G.~K.~Chakravarty and S.~Mohanty, arXiv:1405.1321 [hep-ph]; 
  I.~Oda and T.~Tomoyose, JHEP {\bf 1409}, 165 (2014). 

\bibitem{speedup1}
  J.D. Barrow, Month. Not. R. Astron. Soc. {\bf184}, 677 (1978);
  J. Yang {\it et al.}, Astrophys. J. {\bf 277}, 697 (1979);
  F.S. Accetta {\it et al.}, Phys. Lett. B {\bf248}, 94 (1990).

\bibitem{bbnJFBD}
  J.D. Barrow, Month. Not. R. Astron. Soc. {\bf184}, 677 (1978);
  K. Arai, M. Hashimoto, and T. Fukui, Astron. Asrophys. {\bf179},
  17 (1987);
  F.S. Accetta, L.M. Krauss, and P. Romanelli, Phys. Lett. B
  {\bf248}, 146 (1990);
  T. Damour and C. Gundlach,
  Phys. Rev. D {bf43}, 3873 (1991);
  J.~A.~Casas, J.~Garcia-Bellido and M.~Quiros,
  Mod.\ Phys.\ Lett.\ A {\bf 7}, 447 (1992);
  J.~A.~Casas, J.~Garcia-Bellido and M.~Quiros,
  Phys.\ Lett.\ B {\bf 278}, 94 (1992);
  T. Clifton, J.D. Barrow, and R.J. Scherrer,
 [astro-ph/0504418].

\bibitem{bbnst1}
  A. Serna and J.M. Alimi, Phys. Rev. D {\bf53}, 3074 (1996);
  {\it ibid.}, Phys. Rev. D {\bf53}, 3087 (1996).

\bibitem{bbn_quad}
  D.I. Santiago, D. Kalligas, and R.V. Wagoner,
  {\it ibid.}, Phys. Rev. D {\bf56}, 7627 (1997).

\bibitem{dp}
   T.~Damour and B.~Pichon,
  Phys.\ Rev.\  D {\bf 59}, 123502 (1999).



\bibitem{couv}
  A.~Coc, K.~A.~Olive, J.~P.~Uzan and E.~Vangioni,
  Phys.\ Rev.\  D {\bf 73}, 083525 (2006);
  A.~Coc, K.~A.~Olive, J.~P.~Uzan and E.~Vangioni,
  [arXiv:0811.1845 [astro-ph]].

\bibitem{ru02}
  A. Riazuelo and J.-P. Uzan, Phys. Rev. D {\bf 66}, 023525 (2002);
  A. Riazuelo and J.-P. Uzan, Phys. Rev. D {\bf 62}, 083506 (2000).

\bibitem{carlo}
  C. Schimd, J.-P. Uzan, and A. Riazuelo, Phys. Rev. D {\bf71}, 083512 (2005).

\bibitem{dnord}
   T.~Damour and K.~Nordtvedt,
  Phys.\ Rev.\ Lett.\  {\bf 70}, 2217 (1993);
%
  {\it ibid.}, Phys.\ Rev.\  D {\bf 48}, 3436 (1993).

\bibitem{dpol}
  T.~Damour and A.~M.~Polyakov,
  Nucl.\ Phys.\  B {\bf 423}, 532 (1994).

\bibitem{Khoury:2003aq}
  J.~Khoury and A.~Weltman,
  Phys.\ Rev.\ Lett.\  {\bf 93}, 171104 (2004).

\bibitem{brax}
  P.~Brax, C.~van de Bruck, A.~C.~Davis, J.~Khoury and A.~Weltman,
  Phys.\ Rev.\ D {\bf 70}, 123518 (2004).

\bibitem{HI}
  F.~L.~Bezrukov and M.~Shaposhnikov,  
  Phys.\ Lett.\ B {\bf 659}, 703 (2008); 
  A.~O.~Barvinsky, A.~Y.~Kamenshchik and A.~A.~Starobinsky,  
  JCAP {\bf 0811}, 021 (2008); 
  F.~Bezrukov, D.~Gorbunov and M.~Shaposhnikov,  
  JCAP {\bf 0906}, 029 (2009)
  F.~L.~Bezrukov, A.~Magnin and M.~Shaposhnikov, 
  Phys.\ Lett.\ B {\bf 675}, 88 (2009); 
  F.~Bezrukov,  
  Class.\ Quant.\ Grav.\  {\bf 30}, 214001 (2013). 

\end{references}
\end{document}